\pdfoutput=1

\documentclass{iitpressproc}

\usepackage{graphicx}
\usepackage{amsmath}

\usepackage[sort&compress,merge]{natbib}
\bibpunct{[}{]}{,}{n}{}{}

\begin{document}

\title{Collective dynamics of the p-Pb collisions at the LHC}

\author{Wojciech Broniowski${}^{1,2}$, Piotr Bo\.zek${}^{2,3}$
\address{
         ${}^1$ Institute of Physics, Jan Kochanowski University, 25-406~Kielce, Poland\\
         ${}^2$ The H. Niewodnicza\'nski Institute of Nuclear Physics, Polish Academy of Sciences, 31-342 Krak\'ow, Poland\\
         ${}^3$ AGH University of Science and Technology, Faculty of Physics and Applied Computer Science, 30-059 Krakow, Poland
         }}

\maketitle

\begin{abstract}
We review the signatures for the soft collective dynamics in highest-multiplicity ultrarelativistic p-Pb collisions and show that the effects are well 
described in a three-stage model, consisting of the event-by-event Glauber initial conditions, viscous 3+1D hydrodynamics, and statistical 
hadronization. In particular, we discuss the ridges in two-particle correlations in the relative azimuth and pseudorapidity, the elliptic and triangular flow coefficients, 
and the mass hierarchy of observables sensitive to flow.
\end{abstract}

\section{Introduction}

This talk is based on Refs.~\cite{Bozek:2012gr,Bozek:2013uha,Bozek:2013ska}, where the details and more complete lists of references can be found 
(see also the mini-review~\cite{Bozek:2013yfa}). 
We wish to address here the most intriguing physics questions concerning the topic:

\begin{itemize}

\item Are the highest-multiplicity p-Pb collisions {\em collective?} 
\item What is the nature of the initial state and correlations therein?
\item What are the limits in conditions on applicability of hydrodynamics?

\end{itemize}

Recall that the {\em collective flow} is one of the principal signatures of the strongly-interacting Quark-Gluon Plasma 
formed in ultra-relativistic A-A collisions at RHIC and the LHC. 
It manifests itself in harmonic components in the momentum spectra $v_n$, 
in specific structures in the correlation data {\em (ridges)}, 
in {\em mass hierarchy} of the $p_T$ spectra and $v_n$'s of identified particles, as well as in certain features of interferometry (femtoscopy).  
Since 1)~the ridges were found experimentally at the LHC in p-Pb collisions~\cite{CMS:2012qk,*Abelev:2012ola,*Aad:2012gla,*Aad:2013fja,*Chatrchyan:2013nka}, 
2)~large elliptic and triangular flow was measured in p-Pb~\cite{Aad:2013fja,*Chatrchyan:2013nka},
3)~strong mass hierarchy was recently detected in p-Pb~\cite{Chatrchyan:2013eya,*ABELEV:2013wsa,*Abelev:2013bla},
there are clear analogies between the ``collective'' A-A system and the ``small'' p-A system.
Below we present the evidence for the collective interpretation of the highest-multiplicity p-A collisions.

\section{Three-stage approach}

\begin{figure}
\centering
\includegraphics[width=0.67\columnwidth]{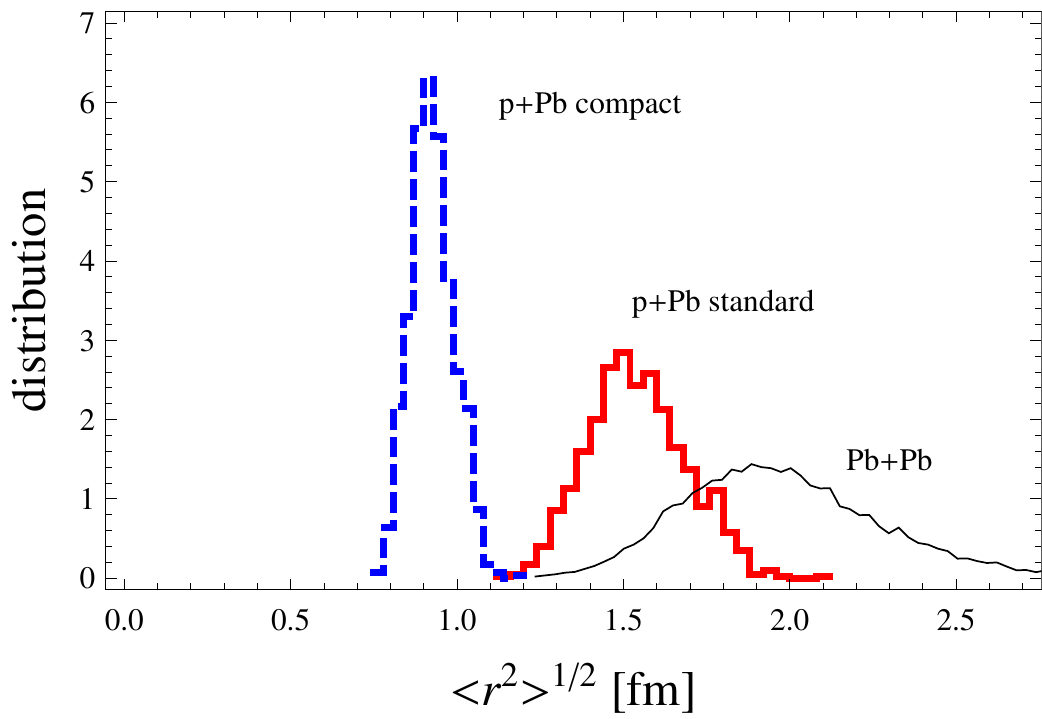} 
\caption{Event-by-event distribution of the rms size of the Glauber initial conditions for the {\em fixed} number of participants, 
$N_p=18$, for the standard source in p-Pb (thick solid line) 
obtained by placing the sources at the centers of participants, the compact source (dashed line), obtained by placing the 
sources in the center-of-mass of the colliding pair~\cite{Bzdak:2013zma}, and for the peripheral Pb-Pb collisions (thin solid line). 
The p-Pb sizes are not more that twice smaller from the Pb-Pb sizes. At the same time, the p-Pb system is more dense. 
This allows to analyze the p-Pb system with viscous hydrodynamics~\cite{Bozek:2013yfa}.
\label{f1}}
\end{figure}

To place our argumentation on a quantitative level, we use the three stage approach consisting of 
1)~ modeling of the initial phase with the Glauber approach as implemented in 
GLISSANDO~\cite{Broniowski:2007nz,*Rybczynski:2013yba}, 2)~applying event-by-event 3+1D viscous 
hydrodynamics~\cite{Bozek:2011ua} to the intermediate evolution, and 3)~carrying out statistical hadronization 
at freezeout with THERMINATOR~\cite{Kisiel:2005hn,*Chojnacki:2011hb}.
The details can be found in Refs.~\cite{Bozek:2012gr,Bozek:2013uha}. Here we only wish to point out the similarity 
of the initial conditions in high-multiplicity p-A collisions to those in peripheral A-A collisions, as seen from Fig.~\ref{f1}. 
This indicates that our approach should work with similar accuracy for the most central p-Pb collisions  
as it worked for the Pb-Pb collisions at centralities~$\sim 70\%$. 

\section{The ridge}

\begin{figure}
\centering
\includegraphics[width=0.85\columnwidth]{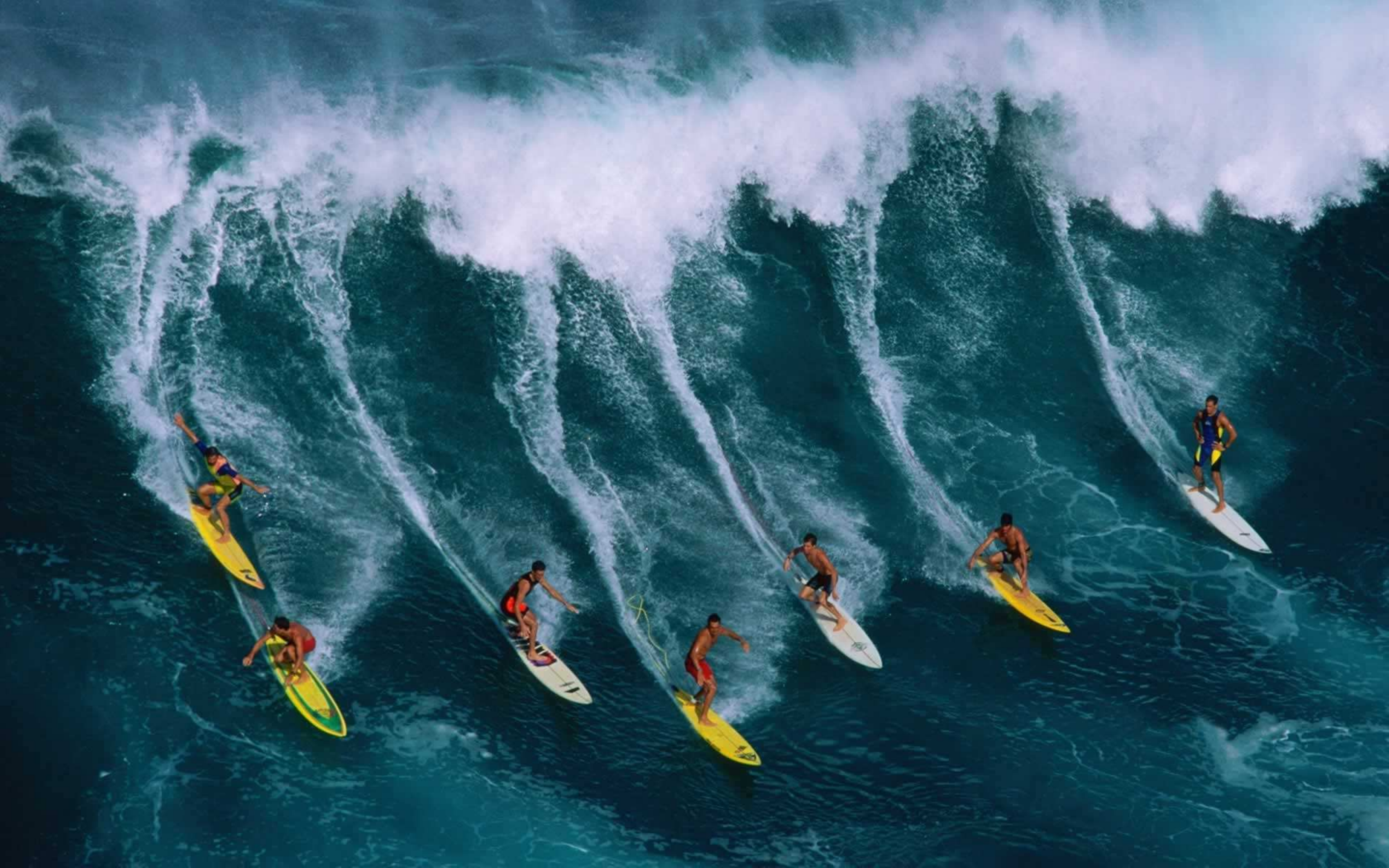}
\caption{Creation of the near-side ridge: 
The surfers' motion is correlated even when they are widely separated along the shore. \label{surf}}
\end{figure}

\begin{figure}
\centering
\includegraphics[width=0.7\textwidth]{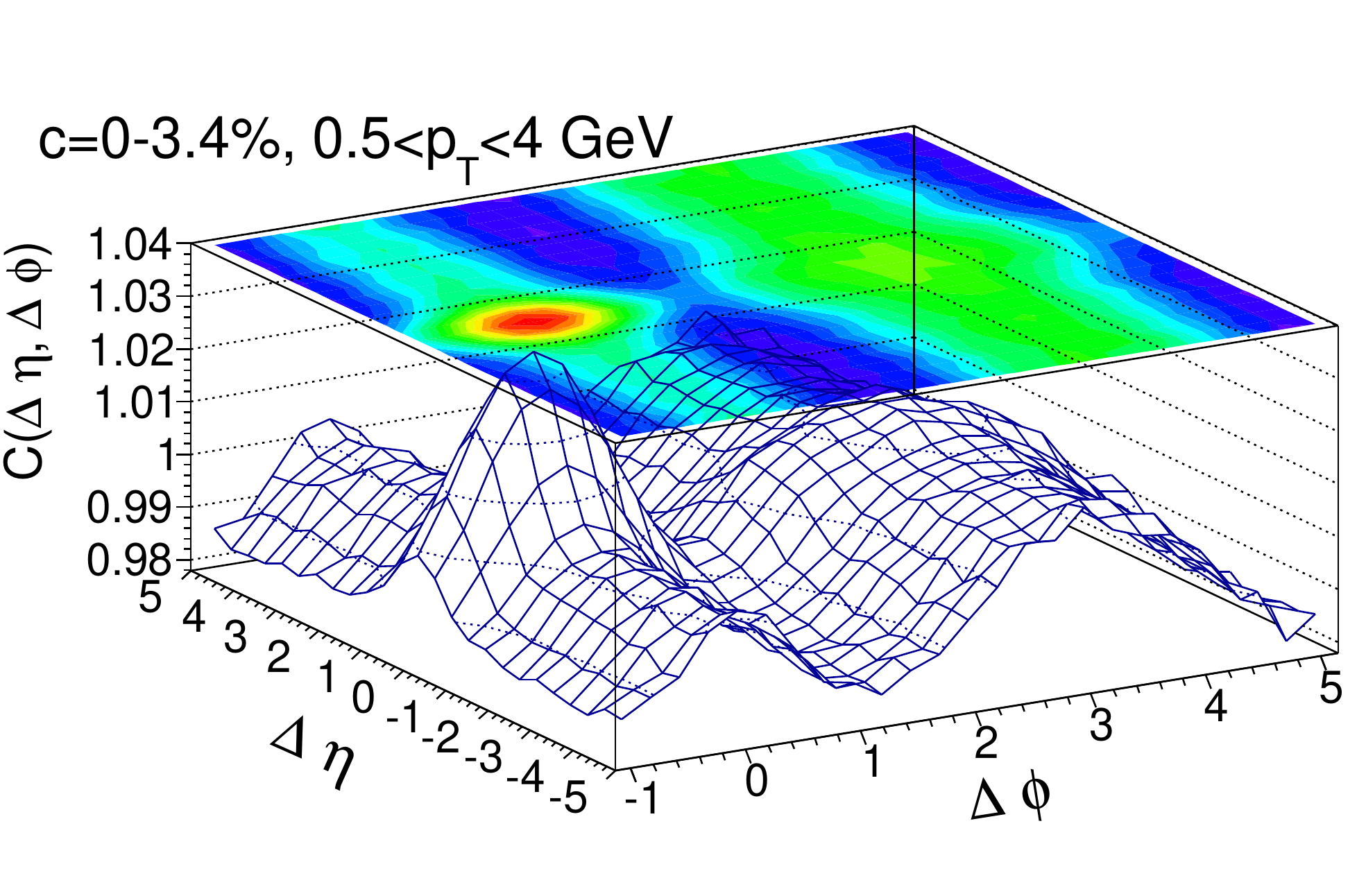} \\ \includegraphics[width=0.6\textwidth]{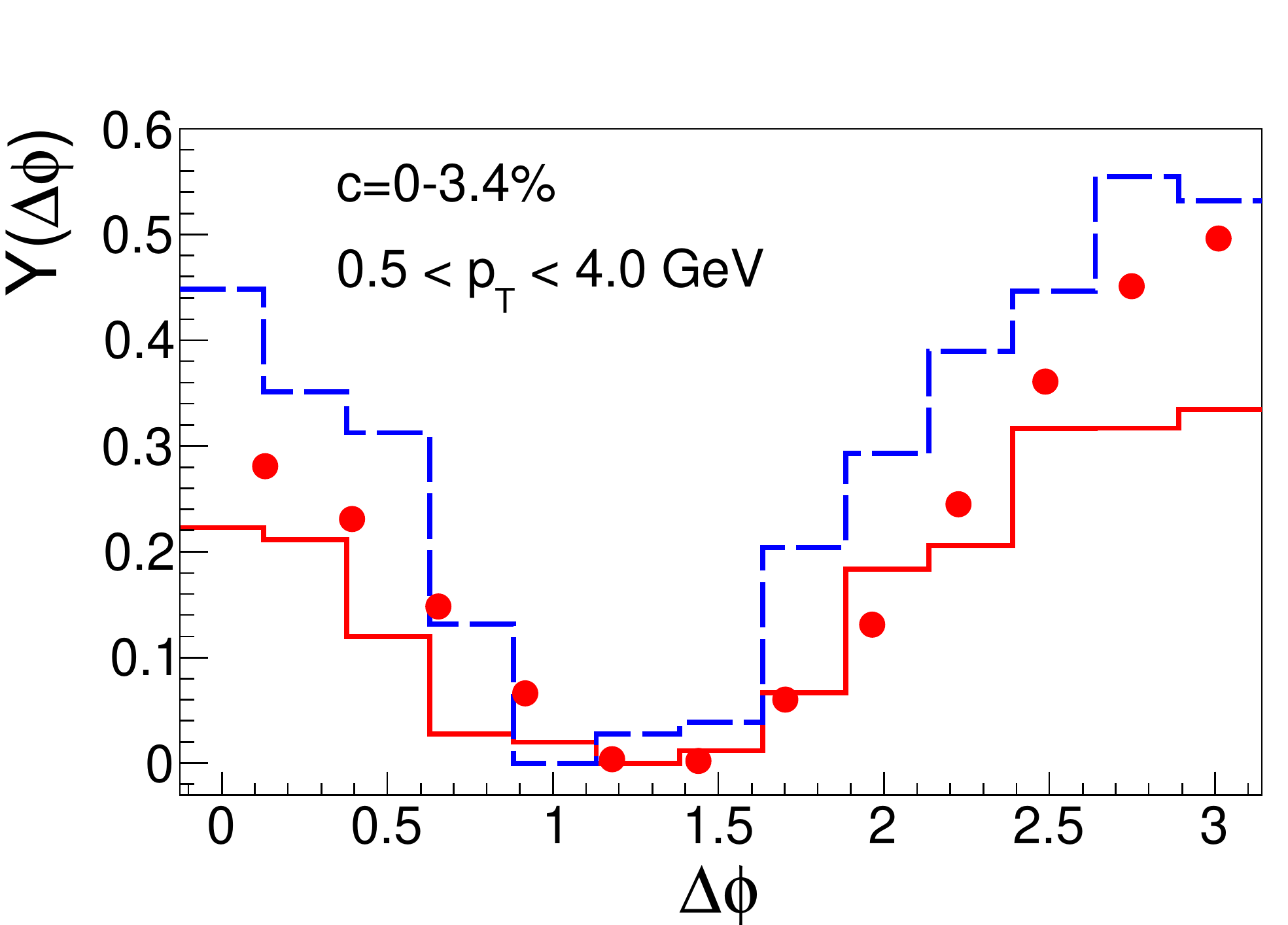} 
\caption{Correlation functions in the ATLAS kinematic conditions. Top:~$C(\Delta \phi,\Delta \eta)$. Bottom: 
the projected ($2\le |\Delta \eta| \le 5$) per-trigger correlation function $Y(\Delta \phi)={\int B(\Delta \phi) d(\Delta \phi)} C(\Delta\phi)/N - b_{\rm ZYAM}$, 
compared to the ATLAS data. The solid (dashed) lines correspond to the standard (compact) source. \label{atlas}}
\end{figure}

The emergence of the ridges in the two-particle correlations in the relative azimuth and pseudorapidity, 
$C(\Delta \eta, \Delta \phi) = {N^{\rm pairs}_{\rm phys}(\Delta \eta, \Delta \phi)}/{N^{\rm pairs}_{\rm mixed}(\Delta \eta)}$, finds a natural 
explanation in correlated collective flow orientation within a long pseudorapidity span. This is cartooned in Fig.~\ref{surf}. Numerical calculation 
in our approach yields fair agreement with the data, as indicated in Fig.~\ref{atlas}, providing alternative explanation to the 
color-glass approach~\cite{Dusling:2012iga,*Dusling:2013oia}. As shown below, the event-by-event 
hydrodynamics also yields the proper magnitude of the triangular component of the flow in a natural way.

\section{Harmonic flow}

\begin{figure}
\vspace{-2cm}
\centering
\includegraphics[width=0.8\textwidth]{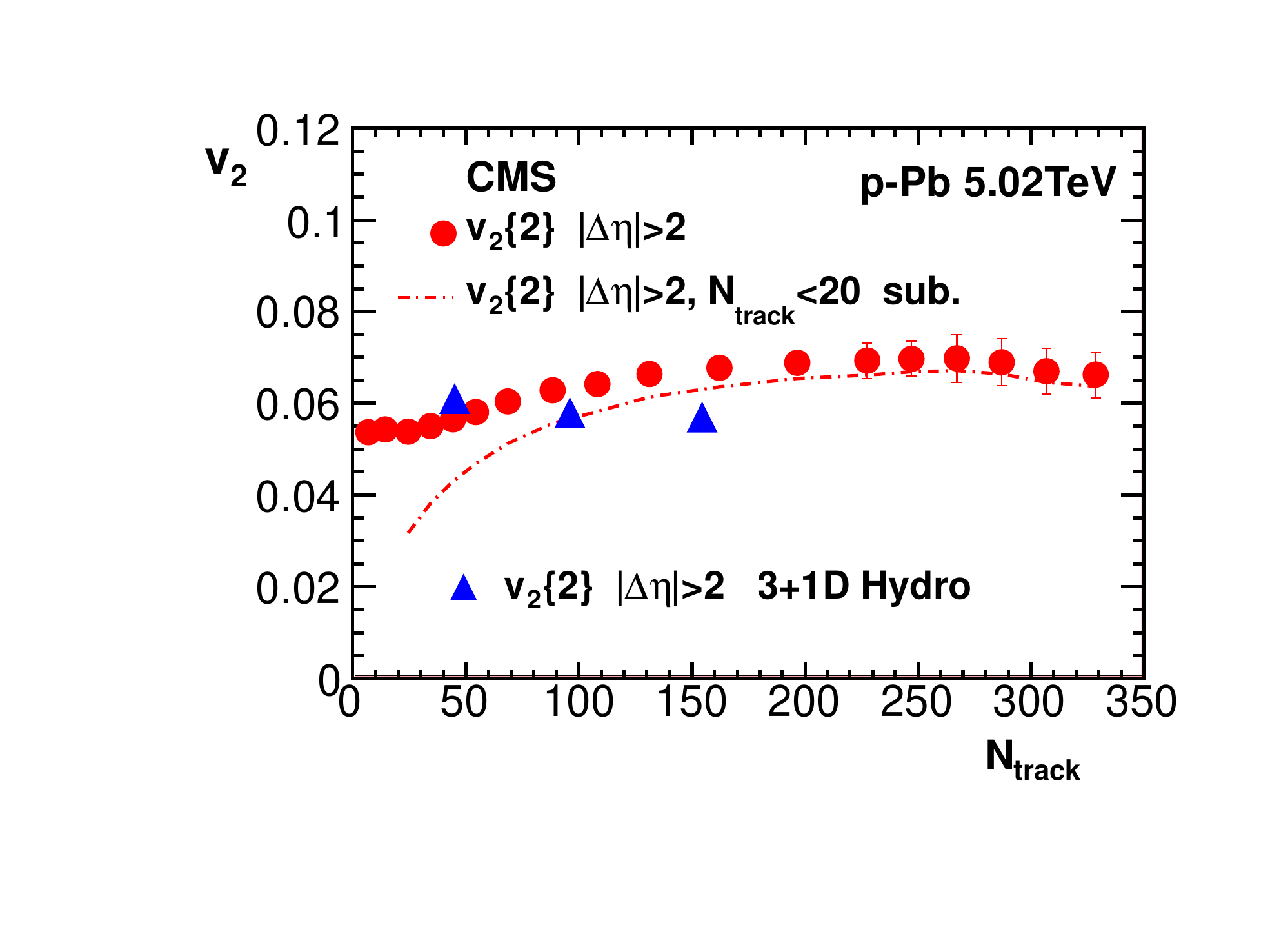} 
\vspace{-15mm}
\caption{The elliptic flow coefficient $v_2\{2, |\Delta \eta >2|\}$ from our model (points) compared to the CMS data. \label{v2cms}}
\end{figure}

\begin{figure}
\vspace{-4mm}
\centering
\includegraphics[width=0.8\textwidth]{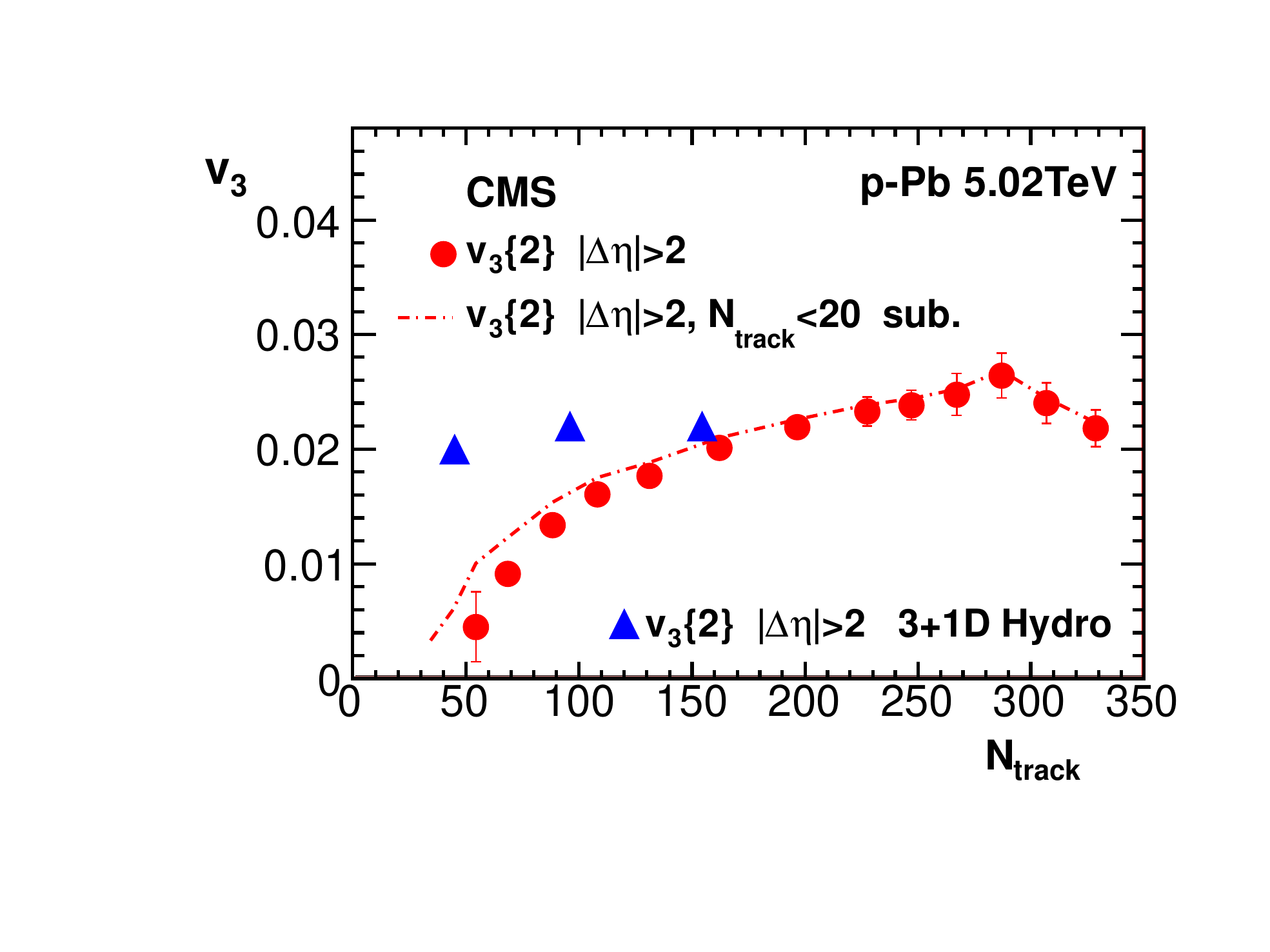} 
\vspace{-15mm}
\caption{Same as Fig.~\ref{v2cms} but for the triangular flow $v_3\{2, |\Delta \eta >2|\}$. The departure of the model from the experiment for lower centrality classes 
indicates the limits of validity of the collective approach. \label{v3cms}}
\end{figure}

The structure of the correlation data (similar to the top panel of Fig.~\ref{atlas})
indicates that one may get rid of most of the nonflow effects by excluding the central peak 
from the analysis, simply using pairs with $|\Delta \eta > 2|$. The flow coefficients 
($v_n\{2, |\Delta \eta >2|\}$) obtained that way from the experiment and from our model simulations are 
compared in Figs.~\ref{v2cms} and \ref{v3cms}. We note a very fair agreement with the data for the 
highest-multiplicity events. As the system becomes smaller, the simulations depart from the experiment, indicating that 
the dissipative effects or the direct production from the corona are becoming important. 

\section{Mass hierarchy}

\begin{figure}
\centering
\includegraphics[width=0.85\textwidth]{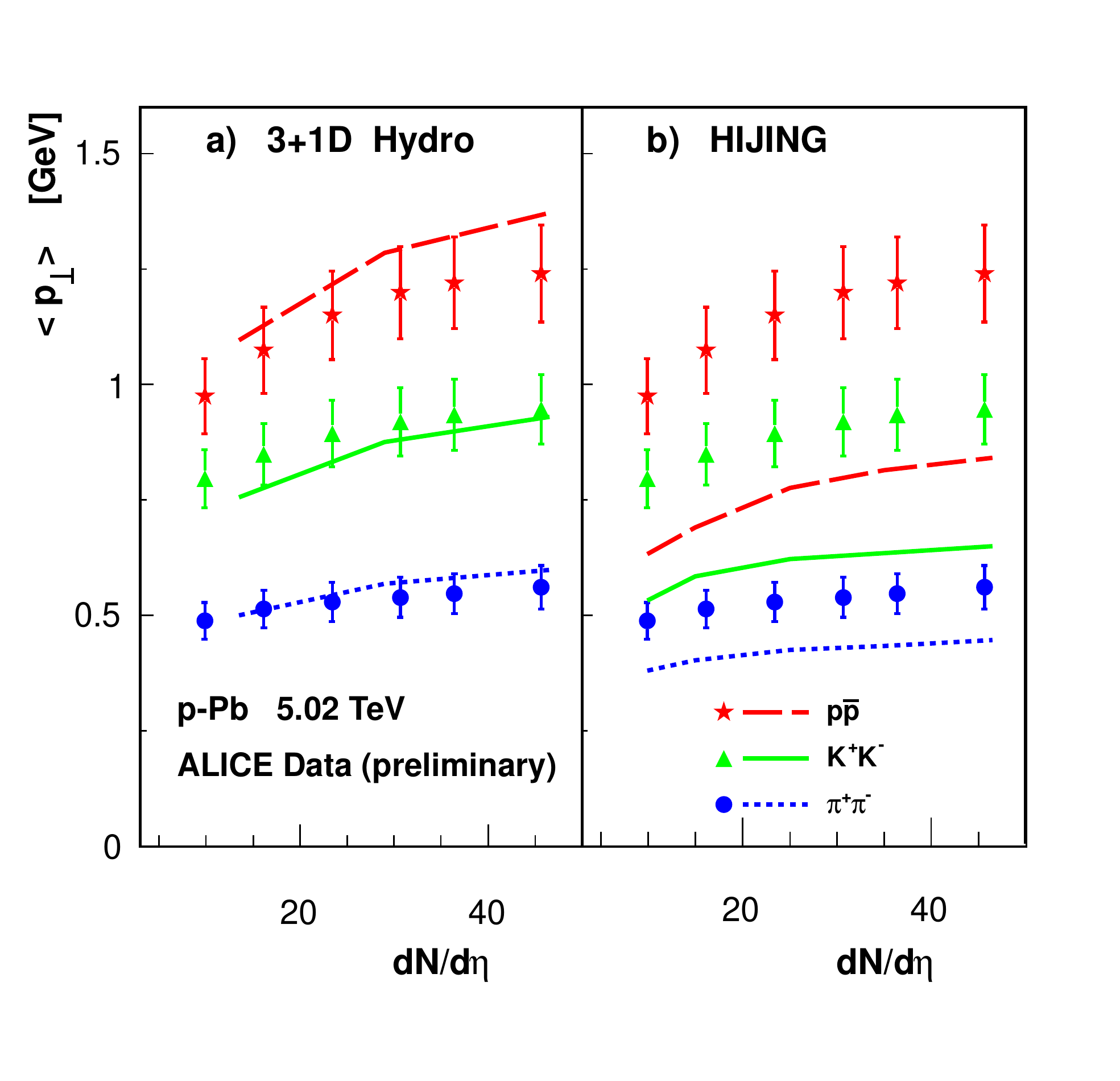} 
\vspace{-9mm}
\caption{Mean transverse momentum of identified particles produced in the p-Pb collisions, plotted as a function of the charged particle density. 
(a)~our model, and for comparison (b)~HIJING~2.1, where no collective effects are present. The lines correspond to the model calculations, while the data points come from Ref.~\cite{Abelev:2013bla}. \label{ids}}
\end{figure}

A very important effect of the presence of collective flow is the emergent mass hierarchy in certain heavy-ion observables~\cite{Bozek:2013ska,Werner:2013ipa}. 
The effect is kinematic: hadrons emitted from a moving fluid element acquire more momentum when they are more massive. 
For that reason, for instance, the average transverse momentum of the protons is significantly higher than for the kaons, which in turn is higher than for the pions. 
The results, showing agreement of our approach with the data, are presented in Fig.~\ref{ids}(a).
As a benchmark with no flow, we show in Fig.~\ref{ids}(b) the results of the HIJING simulations, exhibiting much smaller splitting.

\begin{figure}
\centering
\includegraphics[width=0.7\textwidth]{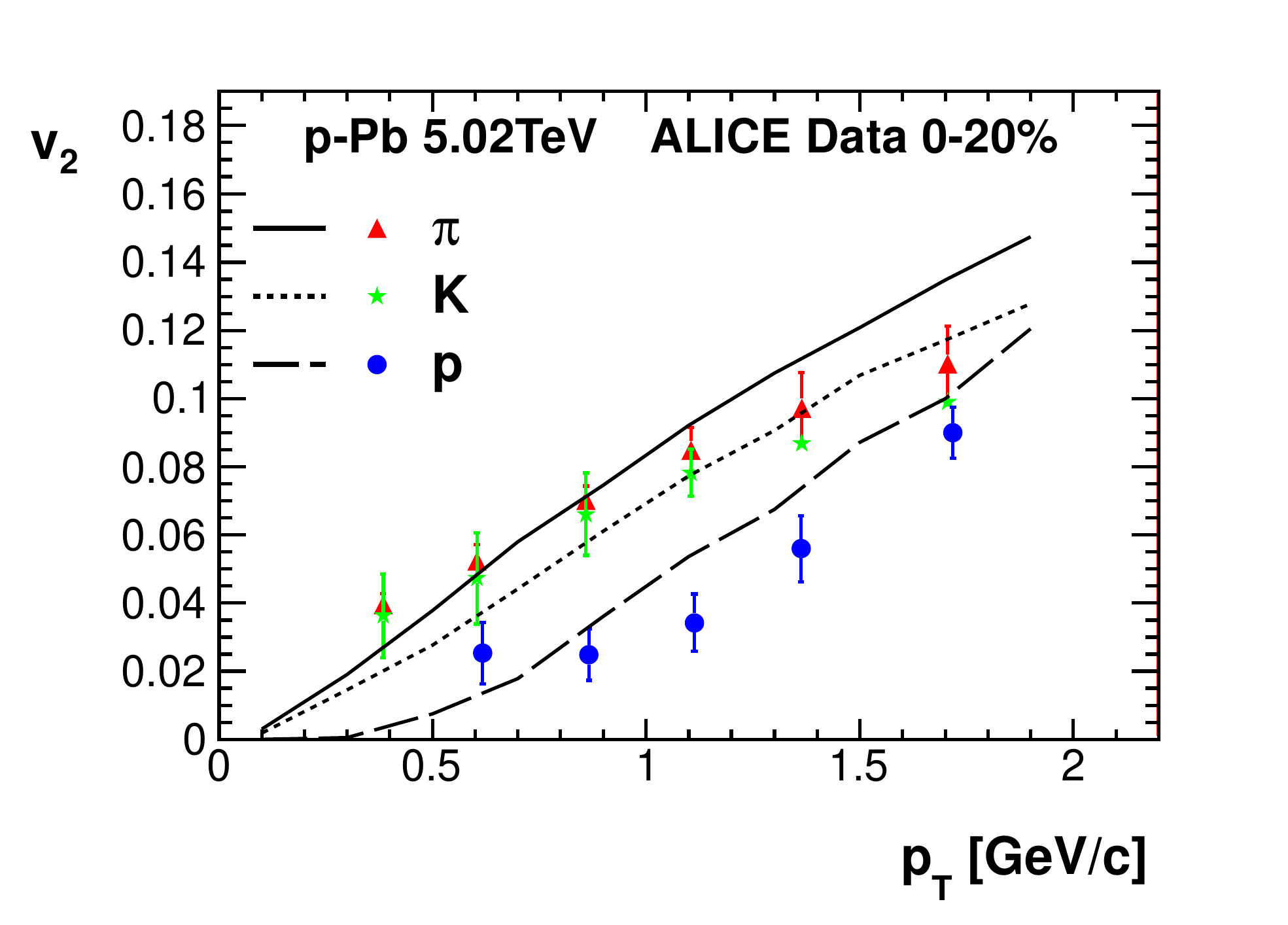} 
\caption{$v_2\{2\}$ for pions, kaons and protons in p-Pb collisions calculated in our model, plotted
as a function of the transverse momentum. The data come from Ref.~\cite{ABELEV:2013wsa}. \label{idv}}
\end{figure}

A proper pattern in the differential
identified-particle elliptic flow is also found, as seen from Fig.~\ref{idv}. We note that a very general argument in favor 
collectivity, based on failure of superposition in the p-A spectra, has been brought up in Ref.~\cite{Bzdak:2013lva}.

\section{Conclusions}

The numerous experimental data from the LHC for the p-Pb collisions of highest multiplicity are 
compatible with the collective expansion scenario: 
the formation of the two ridges, large elliptic and triangular flow, and
the mass hierarchy found in the average transverse momentum and in the differential elliptic flow \cite{Bozek:2013ska,Werner:2013ipa}. 
Thus the p-Pb system can be used as a test ground for the onset of collective dynamics. Certainly, lower multiplicity events are 
``contaminated'' with other effects, e.g., the production from the corona nucleons and their modeling must be more involved.  
Another signature of collectivity
would be provided by the interferometric radii, where the model calculation for p+Pb place the results 
closer to the A+A lines and farther from the p+p lines~\cite{Bozek:2013df}. 

\section*{Acknowledgments}
This work was supported by the Polish National Science Centre, grants DEC-2012/06/A/ST2/00390 and 
DEC-2011/01/D/ST2/00772, and PL-Grid infrastructure.

\bibliographystyle{apsrev}
\bibliography{hydr}

\end{document}